\documentclass[aps,11pt,prd,showpacs,superscriptaddress,preprint]{revtex4}
\usepackage{amsmath}
\usepackage{amssymb}
\usepackage{latexsym}
\usepackage{graphics}
\usepackage{graphicx}
\usepackage{ae,aecompl}

\usepackage{bm}
\usepackage[pdftex,bookmarks=true,bookmarksopen=true,bookmarksnumbered=true,bookmarksopenlevel=3]{hyperref}
\hypersetup{
    bookmarks=true,     
    unicode=false,          
    pdftoolbar=true,        
    pdfmenubar=true,        
    pdffitwindow=false,     
    pdfstartview={FitH},    
    pdftitle={DRAFT},    
    pdfauthor={Orlando Panella},     
    pdfsubject={Dirac equation with position dependent mass},   
    pdfcreator={latex},   
    pdfproducer={TEXSHOP}, 
    pdfkeywords={Dirac Equation, Position dependent mass }, 
    pdfnewwindow=true,      
    colorlinks=false,       
    linkcolor=red,          
    citecolor=green,        
    filecolor=magenta,      
    urlcolor=cyan           
}

\begin{document}

\title{The Casimir Effect in Minimal Length Theories based on a Generalized Uncertainity Principle}

\author{\textsc{A. M. Frassino}}
\affiliation{Frankfurt Institute for Advanced Studies (FIAS), Johann Wolfgang Goethe University, Ruth-Moufang-Strasse 1, Frankfurt am Main, 60438, Germany}

\author{\textsc{O. Panella}}
\affiliation{Istituto Nazionale di Fisica Nucleare, Sezione di Perugia, Via A.~Pascoli, I-06123 Perugia, Italy}

\date{\today}

\begin{abstract}
We study the corrections to the Casimir effect in the classical geometry of two parallel metallic plates, separated by a distance 
$a$, due to the presence of a minimal length ($\hbar\sqrt{\beta}$) arising from  quantum mechanical models based on a 
Generalized Uncertainty Principle (GUP).
 The approach for the quantization of the electromagnetic field is based on projecting onto the maximally localized states of a 
few specific GUP models and was previously developed to study the Casimir-Polder effect. For each model we compute the lowest 
order correction in the minimal length to the Casimir energy and find that it scales with the fifth power of the distance between the 
plates $a^{-5}$  as opposed to the well known QED result which scales  as $a^{-3}$ and, contrary to previous claims, we find that it is 
always attractive. The various GUP models can be in principle differentiated by the strength of the correction to the Casimir energy 
as every model is characterized by a specific multiplicative numerical constant.
\end{abstract}

\pacs{12.90.+b; 12.20.Ds; 12.38.Bx; 04.60.Bc }

\maketitle

\section{Introduction}

The Casimir effect is usually defined as  the physical manifestation of the zero-point energy. It is given by the vacuum fluctuations 
of any quantum field if there are boundary conditions on the field modes. In Casimir's original paper \cite{Casimir:1948dh}, 
the energy is the result of the difference between the vacuum energy of the electromagnetic field in two different configurations: 
the rectangular volume bounded by two parallel conducting plates separated by a distance $a$, along the $\hat{z}$ axis, and 
infinitely extended in the $\left(x,y \right)$ plane, and that of the same volume not bounded by conducting plates. The Casimir 
force is then defined perfoming the usual differentiation of the vacuum energy with respect to the distance $a$ between the plates.
Nevertheless, the Casimir force can be calculated also without references to  the zero-point vacuum fluctuations of quantum fields. In Ref. \cite{PhysRevD.72.021301} the Casimir effect is obtained considering relativistic van der Waals force between the metal plates.  
\\Experimentally this effect can be measured with very high accurancy (for a review, see e.g. \cite{2001PhR...353....1B} and \cite{Milton200110}), but it should be noted that measuring  the Casimir force between two perfectly conducing and parallel plates 
is technically very difficult. Usually, the Casimir force is measured in settings with a plate and a sphere to overcome the problem of parallelism between the plates \cite{PhysRevLett.98.050403}.

The Casimir effect has also been extensively studied from the theoretical point of view because of its connection with the physics beyond the standard model of particle physics. In the literature there are several papers which deal with the corrections to the Casimir energy  due to the existence of a minimal length  (see \cite{Harbach:2005yu}, \cite{2005JPhA...3810027N}), or of compactified extra spatial dimensions \cite{Poppenhaeger:2003es}, or given by a canonical noncommutative spacetime \cite{PhysRevD.76.025016}.\\
The existence of a minimal length in the theory limits explicitly the resolution of small distances in the spacetime. This scale arises naturally in quantum gravity theories in the form of an effective minimal uncertainly in positions $\Delta x_{0}>0$. 
 String theory, for example, predicts  that it is impossible to improve the spatial resolution below the characteristic length of the strings (Refs \cite{PhysRevD.15.2795, string1,Konishi1990276, Michele199365, Michele199383,Michele199383,Garay:1994en}). 
Consequently, these studies yield a correction to the position-momentum uncertainty relation that is related to this characteristic length.
 In one dimension, this minimal length can be implemented adding corrections to the uncertainty relation in this way:
\begin{eqnarray}
\label{MUC}
\Delta x \Delta p \geq \frac{\hbar}{2}\left[1+\beta\left(\Delta p\right)^{2}+\gamma\right],\qquad\beta,\gamma>0 \label{cr}
\end{eqnarray}
which imply the appearance of  a finite minimal uncertainty $\Delta x_{0}=\hbar\sqrt{\beta}$. The development of a generalized quantum theoretical framework which implements the appearance of a nonzero minimal uncertainty in positions is  described in detail in Ref. \cite{kempf-1995-52}. Ref. \cite{2011arXiv1106.2737D} emphasizes that the generalized Eq. \eqref{cr} includes only the first order term of an expansion in the minimum length parameter $\beta$.\\ The modified uncertainty relation Eq. \eqref{cr} implies a small correction term to the usual Heisenberg commutator relation of the form: 
\begin{eqnarray}
\label{MCR}
\left[\hat{x},\hat{p}\right]=i\hbar\left(1+\beta\hat{p}^{2}+\ldots\right).
\end{eqnarray}
Contrary to ordinary quantum mechanics, in these theories the eigenstates of the position operator are no longer physical states whose matrix elements $\langle x|\psi \rangle$ would have the usual direct physical interpretation about positions. One is forced to introduce the  "quasi-position representation", which consists in projecting the states onto the set of maximally localized states. These maximally localized states $| \psi^{ML} _{x} \rangle$ minimize the uncertainty $\left(\Delta x\right)_{|\psi^{ML} _{x} \rangle}=\Delta x_{0}$ and are centered around an average position $\langle  \psi^{ML} _{x} |\hat{x} | \psi^{ML} _{x} \rangle = x$.
In the case of the ordinary commutation and uncertainty relations the  maximally localized states are the usual position eigen-states $| x \rangle$, for which the uncertainty in position vanishes.

In this paper we compute the correction to the Casimir energy arising within a quantum mechanical models based on a Generalized Uncertainty Principle (GUP) given by Eq. \eqref{cr} and generalized to three spatial dimensions.
These results are obtained using the approach developed  in Ref.~\cite{PhysRevD.76.045012} in order to discuss the Casimir-Polder  interaction  within models which include in their theoretical framework a minimal length. 
We will show that quantizing the electromagnetic field as in \cite{PhysRevD.76.045012}, if a minimal length exists in nature the Casimir energy of two large parallel conducting plates, separated by a distance $a$, will acquire, in addiction to the standard $a^{-3}$ interaction, a corrective term which scales as  $a^{-5}$. However, as opposed to previous claims in the literature \cite{2005JPhA...3810027N}, the new term has the same sign of the standard QED result, i.e. it describes an \emph{attractive} interaction. 

The remainder of this paper is organized as follows. In Section II we discuss the generalized uncertainty relations,  introduce a set of maximally localized states and three specific GUP models; at the end of this section we discuss the quantization of the electromagnetic field in the presence of a minimal length; in section III we discuss the standard Casimir effect in QED and, in section IV, we derive the corrections to the Casimir energy due to a minimal length. Finally in sections \ref{discussion} and \ref{conclusion} we present a discussion of our result and our conclusions.

\section{GUP Quantum Mechanics and Second Quantization  }
Let us consider the generalized commutation relations of Eq. \eqref{cr}. In $n$ spatial dimentions  generalized commutation relations which lead to a GUP that provides a minimal uncertainty, assume the form
\begin{eqnarray}
\left[\hat{x}_{i},\,\hat{p}_{j}\right] & = & i\hbar\left[f\left(\hat{p}^{2}\right)\delta_{ij}+g\left(\hat{p}^{2}\right)\hat{p}_{i}\hat{p}_{j}\right]\qquad i,\, j=1,\ldots,n\label{RC}\end{eqnarray}
where the generic functions $f\left(\hat{p}^{2}\right)$ and $g\left(\hat{p}^{2}\right)$ are not completely arbitrary. Relations between them can be found by imposing translational and rotational invariance on the generalized commutation relations.

When the number of dimensions is $n>1$ the generalized uncertainty relations are not unique and different models may be implemented by choosing different functions $f\left(\hat{p}^{2}\right)$ and/or  $g\left(\hat{p}^{2}\right)$ which will yield  different maximally localized states. The specific form of these states depends on the number of dimensions and on the specific model considered. In 
literature there are at least two different approaches to construct maximally localized states: the procedure proposed by Kempf, Mangano and Mann (KMM) \cite{kempf-1995-52} and the one proposed by Detournay, Gabriel and Spindel (DGS) \cite{2002PhRvD..66l5004D}.
The difference lies on the subset of the states to which  the minimization procedure is applied. 
We will see in detail the differences between the results of the two procedures.
As described for Casimir-Polder intermolecular forces \cite{PhysRevD.76.045012}, we analyze two models. The rotationally invariant model (Model I), analyzed adopting both the KMM procedure and the more appropriate DGS method, and the so called direct product model (Model II) used in \cite{2005JPhA...3810027N}.
The general maximally localized states around the average position $\bm{r}$ in the momentum representation can be defined as:
\begin{eqnarray}
\psi_{\bm{r}}^{ML}=\frac{1}{\left(\sqrt{2\pi\hbar}\right)^{3}}\,\Omega\left(p\right)\exp\left\{ -\frac{i}{\hbar}\cdot\left[\bm{\kappa}\left(p\right)\cdot\bm{r}-\hbar\,\omega\left(p\right)t\right]\right\} 
\label{eq:MLS}
\end{eqnarray}
where $p=|\bm{p}|$ and $p^{2}=\bm{p}\cdot\bm{p}=\sum_{i}^{n}\left(p_{i}\right)^{2}$.
The functions $\Omega,\,\bm{\kappa}$ and $\omega$ change for different models. In the following subsections we will review two different models that have been studied in detail in the literature ~\cite{kempf-1995-52,PhysRevD.76.045012}. One of them is studied within two different approaches as regards  the determination of the maximally localized states. We report the the explicit  results of the maximally localized states in the three discussed  examples as these will be used explicitly in the calculations of the Casimir effect in presence of a minimal length. More details can be found in \cite{kempf-1995-52,PhysRevD.76.045012}. 

\subsection{Model I (KMM)}\label{Modello1KMM} This model correspond to  the choice of the generic functions $f\left(\hat{p}^{2}\right)$ and $g\left(\hat{p}^{2}\right)$ given in  Ref. \cite{1997PhRvD..55.7909K}:
\begin{eqnarray}f\left(\hat{p}^{2}\right)=\frac{\beta\hat{p}^{2}}{\sqrt{1+2\beta\hat{p}^{2}}-1},\qquad g\left(\hat{p}^{2}\right)=\beta\label{eq:4}\end{eqnarray}
 From now on we will remove the hat over the operator.
 The KMM construction of maximally localized states gives to Eq.\eqref{eq:MLS} the following functions
\begin{eqnarray}
\kappa_{i}\left(p\right) & = &\left(\frac{\sqrt{1+2\beta p^{2}}-1}{\beta p^{2}}\right)p_{i},\quad\omega\left(p\right)=\frac{pc}{\hbar}\left(\frac{\sqrt{1+2\beta p^{2}}-1}{\beta p^{2}}\right)\quad\label{eq:KMM} \\
\Omega\left(p\right)& = &\left(\frac{\sqrt{1+2\beta p^{2}}-1}{\beta p^{2}}\right)^{\frac{\alpha}{2}}
\label{eq:KMMp}
\end{eqnarray}
where  $n$ is the number of  space dimensions and $\alpha=1+\sqrt{1+n/2}$  is a numerical constant that characterizes the KMM approach. \\
>From the scalar product of maximally localized states   one can define the identity operator:
\begin{eqnarray}
\int\frac{d^{n}p}{\sqrt{1+2\beta p^{2}}}\,\left(\frac{\sqrt{1+2\beta p^{2}}-1}{\beta p^{2}}\right)^{n+\alpha}\left|\bm{p}\right\rangle \left\langle \bm{p}\right|=\bm{1}.\end{eqnarray}

\subsection{Model I (DGS)}\label{Modello1DGS} 
As explained above, different maximally localized states may correspond to a given choice of the generic functions Eq.\eqref{eq:4}. 
The DGS maximally localized states are given by Eq.\eqref{eq:MLS} with:
\begin{eqnarray}
\kappa_{i}\left(p\right) & = & \left(\frac{\sqrt{1+2\beta p^{2}}-1}{\beta p^{2}}\right)p_{i}\qquad\omega\left(p\right)=\frac{pc}{\hbar}\left(\frac{\sqrt{1+2\beta p^{2}}-1}{\beta p^{2}}\right)\label{eq:DGS}\\
\Omega\left(p\right) & = & \left[
\Gamma\left(\frac{3}{2}\right)\left(\frac{2\sqrt{2}}{\pi\sqrt{\beta}}\right)^{\frac{1}{2}}\right]
\left(\frac{1}{p}\frac{\beta p^{2}}{\sqrt{1+2\beta p^{2}}-1}\right)^{\frac{1}{2}}J_{\frac{1}{2}}\left[
\frac{\pi\sqrt{\beta}}{\sqrt{2}}\left(\frac{\sqrt{1+2\beta p^{2}}-1}{\beta p^{2}}\right)p\right]\\
 & = & \frac{\sqrt{2}}{\pi}\frac{\sqrt{\beta p^{2}}}{\left(\sqrt{1+2\,\beta\,{p}^{2}}-1\right)}\sin\left[
 {\frac{\pi\,\left(\sqrt{1+2\,\beta\,{p}^{2}}-1\right)\sqrt{2}}{2\sqrt{\beta\,{p}^{2}}}}\right]\label{eq:DGSp}\end{eqnarray}
 and in this case, the modified identity operator for the momentum eingestates $\left|\bm{p}\right\rangle$ is
 \begin{equation}
\int\frac{d^{n}p}{\sqrt{1+2\beta p^{2}}}\,\left(\frac{\sqrt{1+2\beta p^{2}}-1}{\beta p^{2}}\right)^{n}\left|\bm{p}\right\rangle \left\langle \bm{p}\right|=\bm{1}.\end{equation}

\subsection{Model II}\label{Modello2} 
The model proposed in Ref. \cite{2005JPhA...3810027N} is completely different from that given by Eq. \eqref{eq:4}. This model has 
the functions
\begin{eqnarray}
f\left(p^{2}\right)=1+\beta p^{2},\qquad g\left(p^{2}\right)=0
\end{eqnarray}
in Eq. \eqref{RC},  that give for the maximally localized states
\begin{eqnarray}
\omega\left(p\right)=\frac{c}{\hbar\sqrt{\beta}}\arctan\left(p\sqrt{\beta}\right)\qquad\kappa_{i}\left(p\right)=\left[\frac{1}{\sqrt{\beta}p}\,\arctan\left(p\sqrt{\beta}\right)\right]p_{i}\qquad\Omega\left(p\right)=1\label{eq:MOD2}
\end{eqnarray} 
and the completeness relation reads:
\begin{eqnarray}
\label{eq:MOD22}
\int\frac{d^{3}p}{\left(1+\beta p^{2}\right)}\left|\bm{p}\right\rangle \left\langle \bm{p}\right|=\bm{1}.
\end{eqnarray}

These are the three models that we propose to analyze in this work. We shall now proceed to the quantization of the electromagnetic field following the scheme adopted in Ref. \cite{PhysRevD.76.045012}. In the case of a quantum world with a minimal length the procedure of canonical quantization gets modified: it turns out that the equal-time commutation relations of the fields are different because of the maximally localized states. Instead of expanding the field operators in plane waves (position representation wave functions of momentum states) we are forced to expand the fields in a set of maximally localized states given by Eq. \eqref{eq:MLS}, in this way:
\begin{eqnarray}
\hat{\bm{A}}\left(\bm{r},\, t\right) & = & \sum_{\lambda}\,\int\frac{d^{3}p}{\left(2\pi\right)^{3}}\,\sqrt{\frac{\left(2\pi\right)^{4}\hbar c^{2}}{\omega\left(\bm{p}\right)}}\left[
\hat{a}\left(\bm{p},\,\lambda\right)\varepsilon\left(\bm{p},\,\lambda\right)\langle\psi_{\bm{r}} ^{ML} \right| \left. \bm{p}\rangle+\right. \nonumber \\
& & \left.+\hat{a}^{\dagger}\left(\bm{p},\,\lambda\right)
\varepsilon^{*}\left(\bm{p},\,\lambda\right)
\langle
\bm{p}|
\psi_{\bm{r}} ^{ML}
\rangle 
\right]
\label{PotV}\end{eqnarray}
where $\varepsilon\left(\bm{p},\,\lambda\right)$ are the polarisation vectors. The creation and annihilation operators satisfy the usual commutation relations \[
\left[\hat{a}\left(\bm{p},\,\lambda\right),\,\hat{a}^{\dagger}\left(\bm{p}',\,\lambda'\right)\right]=\left(2\pi\right)^{3}\delta^{\lambda,\lambda'}\bm{\delta}\left(\bm{p}-\bm{p}'\right)\]
and all other commutators vanish.

\section{The Casimir effect in QED}
The Casimir effect  in its simplest form is the interaction of a pair of uncharged, parallel conducting planes caused by the disturbance of the vacuum 
of the electromagnetic field. It is a pure, macroscopic quantum effect because it is only the vacuum, i.e. the ground state of quantum electrodynamics 
(QED), which causes the plates to attract each other. 
Studying the infinite zero-point energy of the quantized electromagnetic field confined between two parallel uncharged plates, 
in his famous paper\cite{Casimir:1948dh}  Casimir derived  the finite energy  between plates.  He found that the energy per unit surface is 
\begin{eqnarray}
\mathcal{E}=-\frac{\pi^{2}}{720}\frac{\hbar c}{a^{3}}\label{eq:2} .  \end{eqnarray}
 where $a$ is the separation between plates along the z-axis, the direction perpendicular to the plates. Consequently the finite force per unit area acting between the plates is $\mathcal{F}=-\frac{\pi^{2}}{240}\frac{\hbar c}{a^{4}}$ and its sign corresponds to an attractive force.
To obtain this result Casimir had renormalized the vacuum energy. He subtracted the \emph{infinite}  vacuum energy of the quantized electromagnetic field in free space (no plates) from the \emph{infinite} vacuum energy  in the presence of plates (at a distance  $a$).  The expression for this energy shift, which turns out to be finite, is  known as the Casimir energy, and reads: 
 \begin{eqnarray}
 \Delta E = \langle 0|\hat{H}\left(a\right)-\hat{H} |0\rangle. \label{shift}
 \end{eqnarray}
For the electromagnetic field in Minkowski space one has to consider the vacuum expectation value of the Hamiltonian operator $\hat{H}$. Choosing the gauge condition $\nabla \cdot \bm{A}=0$ and $\phi = 0$ (see \cite{LandauLifshitz198001}) the Hamiltonian becomes
\begin{eqnarray}
 \hat{H}=\frac{1}{8\pi}\int d^{3}x\,\left[\left(\partial_{0}\hat{\bm{A}}\right)^{2}-\hat{\bm{A}}\nabla^2\hat{\bm{A}}\right]\label{Hamilt} \end{eqnarray}
 that gives
\begin{eqnarray}
E_{0}\equiv \langle 0 |\hat{H}|0\rangle=\frac{1}{2}\hbar\sum_{J}\omega_{J}\label{eq:1}
 \end{eqnarray}
where the index $J$ labels the quantum numbers of the field modes. For the electromagnetic field, the modes are labeled by a three vector $\bm{p}$ in addition to the two polarization $\lambda_i$ $(i=1,2)$ (linear or circular). The indices of the transverse modes indicated collectively as $J$ are thus, in free space (i.e., in the absence of boundaries), the \emph{continuous photon momentum  components} and the two polarization quantum numbers $J=\left(p_{1},\, p_{2},\, p_{3},\lambda_1,\lambda_2\right)$ where all $p_{i} (i=1,3)$ are continuous. After performing the sum over the polarization states the energy in free Minkowski space can be expressed as the integral over a continuous spectrum:
\begin{eqnarray}
\label{energyvac}
E_{0}=\frac{c }{2}\int \frac{L^{2} d^{2}\bm{q}}{\left(2\pi\hbar\right)^{2}} \int^{+\infty}_{-\infty} \frac{a dp_{3}}{\left(2\pi\hbar\right)}\, \sqrt{\bm{q}^2+p_{3} ^2} \label{1}
\end{eqnarray}
with $\bm{q}$ being the transverse momentum in a plane parallel to  the plates.
In the presence of boundaries (the metallic plates) we need to impose the boundary conditions and the result is the quantization of the momentum along the $z$-axis (orthogonal to the plates) $ p_{3} =\frac{n\pi\hbar}{a}$ where  $n=1,2,\ldots$ is the integer quantum number which labels the discrete modes. Thus the  momentum along the $z$ axis is quantized, whereas $\bm{q}=(p_1,p_2)$ takes continuous values. The index of the photon modes becomes now $J=\left(p_{1},\, p_{2},n,\lambda_1, \lambda_2\right)$ and the integral over $dp_{3}$, in the corresponding expression of the energy of Eq.~\eqref{energyvac}, is replaced by a sum over $n$. After summing over the polarization states the energy in the presence of the metallic plates takes the form:
\begin{eqnarray}
E=\frac{c }{2}\int \frac{L^{2} d^{2}\bm{q}}{\left(2\pi\hbar\right)^{2}} \left[|\bm{q}|+2 \sum_{n=1}^{\infty} \sqrt{\left( \bm{q}^{2}+\frac{n^{2} \pi^{2}\hbar^{2}}{a^{2}}\right)}\right].\label{2}\end{eqnarray}
Therefore the energy shift per unit surface is:
\begin{eqnarray}
\mathcal{E} =\frac{c }{\left(2\pi\right)^{2} \hbar^2}\int  d^{2}\bm{q} \left[\frac{1}{2} |\bm{q}|+ \sum_{n=1}^{\infty} \sqrt{\left( \bm{q}^{2}+\frac{n^{2} \pi^{2}\hbar^{2}}{a^{2}}\right)}-\int^{\infty} _{0} dn\, \sqrt{\bm{q}^2+\frac{n^{2} \pi^{2}\hbar^{2}}{a^{2}}}\right] \label{3}\, .
\end{eqnarray}
 Both the equations \eqref{1},\eqref{2} are  ultraviolet divergent for large momenta.
These infinite quantities were regularized using a cutoff function based on the physical reason that, for very short waves,  plates are not an obstacle. 
The equations \eqref{1}, \eqref{2} are then multiplied by some cutoff function of the wave vector $k=|\bm{k}|$, $f\left(k\right)$, such that $f\left(0\right) = 1$ and $f \left( k\gg \frac{1}{a_{0}} \right) \rightarrow 0$, where $a_{0}$ is the typical size of an atom.  Therefore the zero-point energy of these waves will not be influenced by the position of the plates. The presence of the cutoff function  justifies the exchange of sums and integral.
The difference between the sum and the integral, Eq. \eqref{3} becomes:
\begin{eqnarray}
\mathcal{E} =\frac{c }{\left(2\pi\right)^{2} \hbar^2}\left[ \frac{1}{2} G\left( 0 \right) +  G\left( 1 \right) +  G\left( 2 \right) \dots - \int^{\infty} _{0} dn  G\left( n \right)    \right]\, ,
  \label{4}
\end{eqnarray}
with:
\begin{eqnarray}
G\left(n\right) = \int^{+\infty}_{-\infty} d^{2}\bm{q} \sqrt{\bm{q}^2+\frac{n^{2} \pi^{2}\hbar^{2}}{a^{2}}} f\left( \sqrt{\bm{q}^2+\frac{n^{2} \pi^{2}\hbar^{2}}{a^{2}}} \right).
\end{eqnarray}
The difference Eq. \eqref{4}  is evaluated by the Euler-MacLaurin formula, according to which
\begin{eqnarray}
\label{EML}
\nonumber\sum_{n=0}^{N}f\left(n\right)-\int_{0}^{N}dn\, f\left(n\right)& = &-B_{1}\left[f\left(N\right)+f\left(0\right)\right]+\\
& & +\sum_{k=1}^{p}\frac{B_{2k}}{\left(2k\right)!}\left[f^{\left(2k-1\right)}\left(N\right)-f^{\left(2k-1\right)}\left(0\right)\right]+R_p
\end{eqnarray}
where $B_1=-1/2$ is the first Bernoulli's number, $B_{2k}$ are even Bernoulli's numbers, $p$ is an arbitrary integer and $R_p$ is the error term  for the approximation for a given $p$.\\
After subtraction, the regularization was removed leaving the finite result Eq. \eqref{eq:2}.


\section{The Casimir Effect in minimal length QED}
Let us now write the Hamiltonian Eq. \eqref{Hamilt} using the electromagnetic field operator $\hat{\bm{A}}\left(\bm{r},\, t\right)$  decomposed in the so called quasi-position representation~\cite{kempf-1995-52} . The usual plane waves, which are the position representation of   momentum states are replaced by the projection of the momentum states over the complete set of maximally localized states, as in Eq. \eqref{PotV}. 
As we have seen these states are different for each of the above two models and within Model I, for both the KMM and DGS procedure. 
We explicitly derive the Hamiltonian and then the corrections to the Casimir energy due to a minimal length for all models separately. 
In this case we will neglect  surface corrections so that the boundary conditions are defined by: 
\begin{eqnarray} 
\label{bc}
\kappa_3 \left( p \right) = \frac{n\pi\hbar}{a}\,.
\end{eqnarray}

 An important point about the boundary conditions is that in quantum models with a minimal length there is a finite number of modes 
$n_{max}=a/(2\hbar\sqrt{\beta})={a}/{[2\left(\Delta x\right)_{0}]}$. Indeed the  wavelength $\lambda= h/\kappa $ cannot take arbitrary values but has a minimum value  $\lambda_{0}=4\hbar\sqrt{\beta}$. This in turn comes from the fact that, from Eq.~\eqref{eq:MOD2} for example,  one finds that $|\bm{\kappa}|_{max}=\pi/(2\sqrt{\beta}$).   So there is a natural cut-off and the Casimir energy does not need to be regularized,
 as opposed to the standard QED calculation. We start with Model II because in this model both calculation and  notation are simpler due 
to the less complicated functions that describe the maximally localizated states.

\subsection*{Model II} 
We refer to Section~\ref{Modello2}, and use Eqs.~\eqref{eq:MOD2} ,\eqref{eq:MOD22} in Eq.~\eqref{PotV} to write the expansion of the modified  electromagnetic field  $\hat{\bm{A}}\left(\bm{r},\, t\right)$ over the set of maximally localized states, valid in  a quantum theory with a minimal length:
\begin{eqnarray}
\hat{\bm{A}}\left(\bm{r},\, t\right) & = & \sqrt{\frac{c\sqrt{\beta}}{\left(2\pi\right)^{5}\hbar}}\,\sum_{\lambda}\,\int\frac{d^{3}\bm{p}}{\left(1+\beta p^{2}\right)\sqrt{\arctan\left(p\sqrt{\beta}\right)}}\left\{ \left[\varepsilon\left(\bm{p},\,\lambda\right)\hat{a}\left(\bm{p},\,\lambda\right)\, e^{-\frac{i}{\hbar}\left(\kappa\cdot\bm{r}-\hbar\omega t \right)}\right]+\right.\nonumber \\
 &  & \left.+\left[\varepsilon^{*}\left(\bm{p},\,\lambda\right)\hat{a}^{\dagger}\left(\bm{p},\,\lambda\right)\, e^{\frac{i}{\hbar}\left(\kappa\cdot\bm{r}-\hbar\omega t\right)}\right]\right\} \label{eq:0}\end{eqnarray}
The vacuum expectation value of the Hamiltonian operator in free Minkowski space reads:
\begin{eqnarray}
\langle 0|\hat{H}|0\rangle  & = & \frac{1}{8\pi}\,\int d^{3}\bm{r}\,\langle 0|\left(\partial_{0}\hat{\bm{A}}\right)^{2}-\hat{\bm{A}}\nabla^{2}\hat{\bm{A}}|0\rangle \nonumber   =  \frac{1}{2\pi}\,\int d^{3}\bm{r}\int d^{3}\bm{p}\,\frac{c}{\left(2\pi\right)^{2}\hbar^{3}\sqrt{\beta}}\,\frac{\arctan\left(p\sqrt{\beta}\right)}{\left(1+\beta p^{2}\right)^{2}}\nonumber \end{eqnarray}
and explicitly the free space vacuum energy of the quantized electromagnetic field reads:
\begin{eqnarray}
\label{energyp3}
E=\frac{cL^{2}a}{\left(2\pi\right)^{3}\hbar^{3}\sqrt{\beta}}\int_{\mathbb{R}^2} d^{2}\bm{q}\int_{-\infty}^{+\infty} dp_{3}\,\frac{\arctan\left(p\sqrt{\beta}\right)}{\left(1+\beta p^{2}\right)^{2}},
\end{eqnarray}
where $\bm{q}=(p_1,p_2)$ and $p=\sqrt{\bm{q}^2 +p_3^2}$.

In the above equation \eqref{energyp3} we can perform a change of variables from $p_3$ to $\kappa_3$, defined in Model II by   Eq.~\eqref{eq:MOD2}, which will turn out to be more suitable to include the boundary conditions:
\begin{equation}
\label{energykappa3}
E=\frac{cL^{2}a}{\left(2\pi\right)^{3}\hbar^{3}\sqrt{\beta}}\int_{\mathbb{R}^2} d^{2}\bm{q}\int_{-(\kappa_3)_\text{max}}^{+(\kappa_3)_\text{max}} \, d\kappa_3\, \frac{dp_{3}}{d\kappa_3}\,\frac{\arctan\left(p\sqrt{\beta}\right)}{\left(1+\beta p^{2}\right)^{2}},
\end{equation}
where $(\kappa_3)_{\text{max}}= \pi/(2\sqrt{\beta})$.

Indeed the electromagnetic field in the presence of the square parallel plates must satisfy the boundary conditions  as in Eq.~\eqref{bc}, so that $\kappa_{3} (p) ={\hbar\pi n}/{a}$, will give a finite number of discrete values of $\kappa_3$  and $n$ indentifies the \emph{finite number} of modes $n=0,1,2,\ldots n_{max}=(\kappa_3)_\text{max}a /(\hbar\pi)={a}/({2\hbar \sqrt{\beta}})$.  The maximum number of modes $n_\text{max}$ can also be understood by the fact that it is related to the number of minimum wavelengths that it is possible to fit between the plates, $n_\text{max}\,  \lambda_{\text{min} }=  a$.

Now changing again variable from $\kappa_3$ to $n$, the energy shift resulting from the presence of the plates can finally be given  by the relation:
\begin{eqnarray}
\Delta E  & = & \frac{1}{2}\frac{cL^{2}}{\left(2\pi\right)^{2}\hbar^{2}\sqrt{\beta}}\int d^2\bm{q}\,\left\{ \sum_{n=-n_{max}}^{n_{max}}\frac{\arctan\left(
\sqrt{\beta\left[\bm{q}^{2}+p_{3}^{2}\left(n\right)\right]}\right)}{\left\{1+\beta\left[\bm{q}^{2}+p_{3}^{2}\left(n\right)\right]\right\}^{2}}
\left.\frac{d p_3}{d\kappa_3}\right|_{\kappa_3=\frac{\hbar \pi}{a} n}
+\right.\nonumber \\
 &  & \left.-\int_{-n_{max}}^{n_{max}}dn\frac{\arctan\left(\sqrt{\beta\left[\bm{q}^{2}+p_{3}^{2}\left(n\right)\right]}\right)}{\left\{1+\beta\left[\bm{q}^{2}+p_{3}^{2}\left(n\right)\right]\right\}^{2}}\, \left. \frac{d p_3}{d\kappa_3} \right|_{\kappa_3=\frac{\hbar \pi}{a} n}\right\}
\end{eqnarray}
where ${dp_{3}}/{d\kappa_3}$ is a function of $p_3(n)$ through Eqs.~\eqref{eq:MOD2},\eqref{bc} . The energy shift per unit area is: 
 \begin{equation}\mathcal{E}=\frac{\Delta E}{L^{2}}= \frac{c}{\left(2\pi\right)^{2}\hbar^{2}}\,\left\{ \frac{1}{2}G\left(0\right)+\sum_{n=1}^{n_{max}}G\left(n\right)-\int_{0}^{n_{max}}dn\, G\left(n\right)\right\} \label{eq:Shift}
\end{equation}
where, exchanging sums and integrals, we have defined:
\begin{equation}
G\left(n\right)=\frac{1}{\sqrt{\beta}}\int_{-\infty}^{\infty}d\bm{q}\frac{\arctan\left(\sqrt{\beta\left[\bm{q}^{2}+p^{2}_{3}\left(n\right)\right]}\right)}{\left\{1+\beta\left[\bm{q}^{2}+p_{3}^{2}\left(n\right)\right]\right\}^{2}} \, \left.\frac{dp_{3}}{d\kappa_3}\right|_{\kappa_3=\frac{\hbar \pi}{a} n}
.\label{eq:Int}
\end{equation}
We would like to emphasize here that in contrast to the usual QED calculation of the Casimir energy, in this case the expression for the Casimir energy density is finite and does not need to be regularized. The sum in Eq. \eqref{eq:Shift} is over a finite number of terms, and for the same reason the integral converges.  
The minimal length $\hbar \sqrt{\beta}$ induces a natural cut-off. \\
Performing an appropriate change of variables we obtain:
\begin{eqnarray}
\label{G(n)}
G\left(n\right)  & = & \frac{1}{\sqrt{\beta}}\int_{0}^{2\pi}d\theta\int_{0}^{\infty}r\, dr\,\frac{\arctan\left(\sqrt{\beta\left[r^{2}+p^{2}_{3}\left(n\right)\right]}\right)}{\left\{1+\beta\left[r^{2}+p_{3}^{2}\left(n\right)\right]\right\}^{2}}\, \left. \frac{dp_{3}}{d\kappa_3}\right|_{\kappa_3=\frac{\hbar \pi}{a} n}\nonumber\\
& = & \frac{\pi}{\sqrt{\beta}}\int_{0}^{\infty}dx\,\frac{\arctan\left(\sqrt{\beta\left[x+p^{2}_{3}\left(n\right)\right]}\right)}{\left\{1+\beta\left[x+p_{3}^{2}\left(n\right)\right]\right\}^{2}}\, \left.\frac{dp_{3}}{d\kappa_3}\right|_{\kappa_3=\frac{\hbar \pi}{a} n}\,.
\end{eqnarray}
As we have already pointed out the minimal length introduces a natural cut-off through the finiteness of $n_{max}$.  In order to extract the finite result from Eq.~\eqref{eq:Shift} we would need to compute numerically $p_3(n)$ inverting Eq.~\eqref{bc} and then evaluate the factor ${dp_{3}}/{d\kappa_3}|_{\kappa_3={\hbar \pi n}/{a} }$. Finally one has to compute, again numerically, the integral in Eq.~\eqref{G(n)} which 
defines $G(n)$.

We decided however to deduce analytically the  lowest order correction in $\beta$ to the Casimir energy.  We can perform the integral introducing  the simplifying assumption that $p_3(n)= \hbar\pi n/a$, valid in the limit of $\beta \to 0$  or inverting Eq.~\eqref{bc} for $p_3(n)$ via Eq.~\eqref{eq:MOD2} (again in the limit $\beta \to 0$). In the same leading order in $\beta$ we have of course also: ${dp_{3}}/{d\kappa_3}|_{\kappa_3={\hbar \pi n}/{a} } \to 1$. We obtain then a closed expression for $G(n)$:
 \begin{eqnarray}
G\left(n\right) & = & -\frac{\pi}{4} {{\beta}^{-3}\left({\frac{\beta\,{\hbar}^{2}{n}^{2}{\pi}^{2}}{{a}^{2}}}+1\right)}^{-1} 
\left(-2\arctan\left({\frac{\sqrt{\beta}\hbar\pi n}{a}}\right){\beta}^{3/2}+\right.\nonumber\\
& &\left.+2\,{\beta}^{5/2}\arctan\left({\frac{\sqrt{\beta}\hbar\pi n}{a}}\right)\hbar^{2}{\pi}^{2}{n}^{2}{a}^{-2}+2{\frac{\pi\hbar n{\beta}^{2}}{a}}-{\frac{{\pi}^{3}{\beta}^{5/2}\hbar^{2}{n}^{2}}{{a}^{2}}}-\pi{\beta}^{3/2}\right)
.\label{eq:Gn}\end{eqnarray}
Now we can extend the sum and the integral in Eq. \eqref{eq:Shift} to infinity. In fact, $n_{max}$, which is given by the presence in the model  of the minimum wavelength $\lambda_{0}=4\hbar\sqrt{\beta}$, is by definition the greatest integer for  which the transverse modes satisfy the boundary conditions in Eq.~\eqref{bc}, and $n_{max}= a/(2\hbar \sqrt{\beta}) = a/(\lambda_{min}/2) \to \infty $ as $\beta\to 0$.  So we can approximate $n_{max}\rightarrow \infty$ and then we can apply the 
Euler-MacLaurin formula of Eq.~\eqref{EML} with $N= \infty$  and all contributions of the function $G(n)$ and of its derivatives vanishing at infinity (as we have explicitly verified):
\begin{equation}
\frac{1}{2}G\left(0\right)+ \sum_{n=1} ^{\infty} G\left( n \right)
-\int_{0}^{\infty}dn\, G\left(n\right)
=-\frac{1}{2!}B_{2}G'\left(0\right)-\frac{1}{4!}B_{4}G'''\left(0\right)-\frac{1}{6!}B_{6}G^{\mathsf{v}}\left(0\right)+\ldots\label{eq:EM}
\end{equation}
The limiting values of the derivatives of $G\left( n\right)$ at $n=0$ are easily computed to:
\begin{eqnarray}
\lim_{n\rightarrow0}G'\left(n\right) & = & 0;\\
\lim_{n\rightarrow0}G'''\left(n\right) & = &-4\,{\frac{{\hbar}^{3}{\pi}^{4}}{{a}^{3}}};\\
\lim_{n\rightarrow0}G^{\mathsf{v}}\left(n\right) & = &112\,{\frac{{\pi}^{6}{\hbar}^{5}\beta}{{a}^{5}}}.
\end{eqnarray}
One obtains then the  final result with the first order correction term in the minimal uncertainty parameter $\beta$ introduced in the modified  commutation relations of Eq.~\eqref{MCR} : 
\begin{eqnarray}
\mathcal{E} 
  =  -\frac{\pi^{2}}{720}\frac{\hbar c}{a^{3}}\left[1+\pi^{2}\frac{2}{3}\left(\frac{\hbar\sqrt{\beta}}{a}\right)^{2}\right].\label{eq:Risultato}
\end{eqnarray}
The first term in equation Eq. \eqref{eq:Risultato} is the usual Casimir energy reported in Eq. \eqref{eq:2} and is obtained without the cut-off function. 
The second term is the correction given by the presence in the theory of a minimal length. We note that it is attractive.
The Casimir pressure between the plates is given by $\mathcal{F}=-\frac{\partial}{\partial a}\mathcal{E}$
\begin{eqnarray}
\mathcal{F}= -\frac{\pi^{2}}{240}\frac{\hbar c}{a^{4}}\left[1+\pi^{2}\frac{10}{9}\left(\frac{\hbar\sqrt{\beta}}{a}\right)^{2}\right].    
\end{eqnarray}



\subsection*{Model I (KMM)} 
Here we refer to Section \ref{Modello1KMM}, 
and replacing Eqs. \eqref{eq:KMM},\eqref{eq:KMMp} in the definition of the modified electromagnetic  field as given in Eq.~\eqref{PotV},  and calculating the vacuum expectation value of the Hamiltonian, we obtain:
\begin{eqnarray}
\langle 0|\hat{H}|0\rangle & = & \frac{1}{8\pi}\,\int d^{3}\bm{r}\int d^{3}\bm{p}\,\frac{4c}{\left(2\pi\right)^{2}\hbar^{3}}\,\frac{\left(\sqrt{1+2\beta p^{2}}-1\right)^{7+3\alpha}}{\beta^{7+3\alpha}p^{2(\frac{13}{2}+3\alpha)}\left(1+2\beta p^{2}\right)}.\nonumber 
\end{eqnarray}
 The energy shift is: 
 \begin{eqnarray}
\Delta E  & = & \frac{1}{2}\frac{cL^{2}}{\left(2\pi\right)^{2}\hbar^{2}}\int d^2\bm{q}\,\left\{ \sum_{n=-n_{max}}^{n_{max}}\frac{\left(\sqrt{1+2\beta\left[\bm{q}^{2}+p_{3}^{2}\left(n\right)\right]}-1\right)^{7+3\alpha}}{\beta^{7+3\alpha}\left[\bm{q}^{2}+p_{3}^{2}\left(n\right)\right]^{(\frac{13}{2}+3\alpha)}\left\{1+2\beta\left[\bm{q}^{2}+p_{3}^{2}\left(n\right)\right]\right\}} \, \left.\frac{dp_{3}}{d\kappa_3}\right|_{\kappa_3=\frac{\hbar \pi}{a} n}\,\right.\nonumber \\
 &  & \left.-\int_{-n_{max}}^{n_{max}}dn\frac{\left(\sqrt{1+2\beta\left[\bm{q}^{2}+p_{3}^{2}\left(n\right)\right]}-1\right)^{7+3\alpha}}{\beta^{7+3\alpha}\left[\bm{q}^{2}+p_{3}^{2}\left(n\right)\right]^{(\frac{13}{2}+3\alpha)}\left\{1+2\beta\left[\bm{q}^{2}+p_{3}^{2}\left(n\right)\right]\right\}} \, \left.\frac{dp_{3}}{d\kappa_3}\right|_{\kappa_3=\frac{\hbar \pi}{a} n}\, \right\} 
\end{eqnarray} 
where $n_{max}={a}/({2\hbar \sqrt{\beta}})$. Interchanging sums and integrals it is possible to define:
\begin{equation}
G_{KMM}\left(n\right)=\int_{-\infty}^{\infty}d^2\bm{q}\frac{\left(\sqrt{1+2\beta\left[\bm{q}^{2}+p_{3}^{2}\left(n\right)\right]}-1\right)^{7+3\alpha}}{\beta^{7+3\alpha}\left[\bm{q}^{2}+p_{3}^{2}\left(n\right)\right]^{(\frac{13}{2}+3\alpha)}\left\{1+2\beta\left[\bm{q}^{2}+p_{3}^{2}\left(n\right)\right]\right\}} \, \left.\frac{dp_{3}}{d\kappa_3}\right|_{\kappa_3=\frac{\hbar \pi}{a} n}\, .\label{G_nkmm}
\end{equation}
Here $p_{3} \left( n \right)$ would indicate the solution to  the boundary conditions in Eq.~\eqref{bc}  but now through Eq.~\eqref{eq:MLS}. Again we are interested in deducing the first order in $\beta$ corrective term to the Casimir energy and therefore we compute the function $p_{3}(n)$ in the limit of $\beta\to 0$ and thus we have $p_3(n) \to n\pi\hbar / a$ and $dp_3/d\kappa_3 \to 1$.
Differently than in the case of Model II here it is not possible to compute the integral in Eq. \eqref{G_nkmm} in closed form. One can find however its derivatives with respect to $n$ :
\[
G'_{KMM}\left(n\right)=-2\frac{\left(2\right)^{(\frac{13}{2}+3\alpha)}}{\sqrt{\beta}}{\hbar}^{2}{\pi}^{4}n\sqrt{2}\sqrt{{\frac{\beta\,{\hbar}^{2}{n}^{2}}{{\mbox{a}}^{2}}}}\left(\sqrt{{\frac{{a}^{2}+2\beta{\hbar}^{2}{\pi}^{2}{n}^{2}}{{a}^{2}}}}+1\right)^{-7-3\,\alpha}\left({a}^{2}+2\beta{\hbar}^{2}{\pi}^{2}{n}^{2}\right)^{-1}\]
and substituting $\alpha=1+\sqrt{1+\frac{3}{2}}$ we find:
\begin{eqnarray}
\lim_{n\rightarrow 0^+}G'_{KMM}\left(n\right) & = & 0\, ;\\
\lim_{n\rightarrow 0^{+}}G'''_{KMM}\left(n\right)
 & = & -4\,{\frac{{\hbar}^{3}{\pi}^{4}}{{a}^{3}}}\, ;\\
 \lim_{n\rightarrow0^{+}}G^{\mathsf{\,v}}_{KMM} \left(n\right)
 & = & \left(336+36\sqrt{10}\right)\frac{{\pi}^{6}{\mbox{\ensuremath{\hbar}}}^{5}{\beta}}{{a}^{5}}\, .\end{eqnarray}
 We did check that at $N=n_{max} \to \infty$ (in the limit of $\beta\to 0$)  all derivatives of the function $G_{KMM}(n)$ vanish  and do not contribute to the Euler-MacLaurin formula of Eq.~\eqref{EML}. 
We can thus apply again the Euler-MacLaurin formula as given in  Eq.~\eqref{eq:EM} to calculate the Casimir energy, and we finally get:
\begin{eqnarray}
 \mathcal{E} & = & -\frac{\pi^{2}}{720}\frac{\hbar c}{a^{3}}\left[1+\pi^{2}\left(\frac{28+3\sqrt{10}}{14}\right)\left(\frac{\hbar\sqrt{\beta}}{a}\right)^{2}\right]. \label{eq:RisultatoK}\end{eqnarray}
We note that the first order correction in $\beta$ turns out to describe an attractive interaction. The Casimir pressure is:
\begin{eqnarray}
\mathcal{F}= -\frac{\pi^{2}}{240}\frac{\hbar c}{a^{4}}\left[1+\pi^{2}\left( \frac{10}{3}+\frac{5\sqrt{10}}{14}\right) \left(\frac{\hbar\sqrt{\beta}}{a}\right)^{2}\right].
\end{eqnarray}


\subsection*{Model I (DGS)} 
This time we refer to Section \ref{Modello1DGS}. 
The modified potential vector describing the modified electromagnetic field   $\hat{\bm{A}}\left(\bm{r},\, t\right) $ obtained substituing Eq. \eqref{eq:DGS} ,\eqref{eq:DGSp} in the definition Eq. \eqref{PotV}  leads by a similar procedure to the vacuum energy: 
\begin{eqnarray} 
E & = & \frac{c L^2 a}{\left(2\pi\right)^{3}\hbar^{3}}\int d^{3}\bm{p}\,\left(\frac{2}{\pi^{2}}\right)\frac{\left(\sqrt{1+2\beta p^{2}}-1\right)^{5}\sin^{2}\left[{\frac{\pi\,\left(\sqrt{1+2\beta p^{2}}-1\right)}{\sqrt{2\beta}p}}\right]}{{\beta}^{6}\left(p^{2}\right)^{\frac{11}{2}}}\label{sss3}
\end{eqnarray} 
then the energy shift is explicitly given by:
\begin{eqnarray}
\Delta E 
 & = & \frac{1}{2}\frac{cL^{2}}{\left(2\pi\right)^{2}\hbar^{2}}\left(\frac{2}{\pi^{2}}\right)\int d^2\bm{q}\,\left\{ \sum_{n=-n_{max}}^{n_{max}}\frac{\left(\upsilon-1\right)^{5}\sin^{2}\left[{\frac{\pi\,\left(\upsilon-1\right)}{\sqrt{\upsilon^2 - 1}}}\right]}{{\beta}^{6}\left[\bm{q}^{2}+p_{3}^{2}\left(n\right)\right]^{\frac{11}{2}}\upsilon^2} \, \left.\frac{dp_{3}}{d\kappa_3}\right|_{\kappa_3=\frac{\hbar \pi}{a} n}\,\right.\nonumber \\
 &  & \left.-\int_{-n_{max}}^{n_{max}}dn\frac{\left(\upsilon-1\right)^{5}\sin^{2}\left[{\frac{\pi\,\left(\upsilon-1\right)}{\sqrt{\upsilon^2 -1}}}\right]}{{\beta}^{6}\left[\bm{q}^{2}+p_{3}^{2}\left(n\right)\right]^{\frac{11}{2}}\upsilon^2}\, \left.\frac{dp_{3}}{d\kappa_3}\right|_{\kappa_3=\frac{\hbar \pi}{a} n}\,\right\}. \end{eqnarray} 
 where $\upsilon=\sqrt{1+2\beta\left[\bm{q}^{2}+p_{3}^{2}\left(n\right)\right]}$  and $\bm{q}=(p_1,p_3)$.\\
Again interchanging  sums and integral, we can define the  function:
\begin{equation}
\label{GDGS(n)}
G_{DGS}\left(n\right)=\left(\frac{2}{\pi^{2}}\right)\int_{-\infty}^{\infty}d^2\bm{q}\frac{\left(\upsilon-1\right)^{5}\sin^{2}\left[{\frac{\pi\,\left(\upsilon-1\right)}{\sqrt{\upsilon^2 -1}}}\right]}{{\beta}^{6}\left[\bm{q}^{2}+p_{3}^{2}\left(n\right)\right]^{\frac{11}{2}}\upsilon^2}\,  \, \left.\frac{dp_{3}}{d\kappa_3}\right|_{\kappa_3=\frac{\hbar \pi}{a} n}\,.
\end{equation}
From here we proceed as we did in Model II and in Model I(KMM) and use the  $p_{3} \left( n \right)$ which satisfies the boundary conditions in Eq.~\eqref{bc}  but now through Eq.~\eqref{eq:DGS} . In order to deduce the first order in $\beta$ corrective term to the casimir energy  we compute the function $p_{3}(n)$ in the limiting case $\beta\to 0$ and thus we have $p_3(n)  \to n\pi\hbar / a$ and $dp_3/d\kappa_3 \to 1$.
Again it is not possible to compute the integral in Eq. ~\eqref{GDGS(n)} in closed form. 

One can compute however the derivative of $G_{DGS}(n)$ with respect to $n$  :
\begin{eqnarray}
G'_{DGS}\left(n\right)=-\frac{2\left(2\right)^{\frac{11}{2}}}{\pi\beta\sqrt{\beta}}\cdot\frac{1}{32}\,\hbar^{2}{\pi}^{2}n\sqrt{2}\frac{\left(\sqrt{1+2\,{\frac{\beta\,{\mbox{\ensuremath{\hbar}}}^{2}{\pi}^{2}{n}^{2}}{{a}^{2}}}}-1\right)^{5}\sin^{2} \left[\frac{\pi\,\left(\sqrt{1+2\,{\frac{\beta\,{\hbar}^{2}{\pi}^{2}{n}^{2}}{{a}^{2}}}}-1\right)}{\sqrt{2\beta{\frac{{\hbar}^{2}{\pi}^{2}{n}^{2}}{{a}^{2}}}}}\right]}{\left(1+2\,{\frac{\beta\,{\hbar}^{2}{\pi}^{2}{n}^{2}}{{a}^{2}}}\right){\beta}^{\frac{9}{2}}{a}^{2}\left({\frac{{\hbar}^{2}{\pi}^{2}{n}^{2}}{{a}^{2}}}\right)^{\frac{11}{2}}}\, .
\end{eqnarray}

Now assuming as in the previous examples the limit $\beta\to 0$ , we can compute the first contributing terms (in $n=0$) to  the Eulero-MacLaurin formula in Eq.~\eqref{eq:EM}:
\begin{eqnarray}
\lim_{n\rightarrow 0^{+}}G'_{DGS}\left(n\right) & = & 0\\\lim_{n\rightarrow0^{+}}G'''_{DGS}\left(n\right) & = & -4\,{\frac{{\hbar}^{3}{\pi}^{4}}{{a}^{3}}}\\ \lim_{n\rightarrow0^{+}}G^{\mathsf{\,v}}_{DGS} \left(n\right) & = & {\frac{{\pi}^{6}{\mbox{\ensuremath{\hbar}}}^{5}{\beta}}{{a}^{5}}\cdot8\left(33+{\pi}^{2}\right)}
\end{eqnarray}

 We did check that at $N=n_{max} \to \infty$ (in the limit of $\beta\to 0$)  all derivatives of the function $G_{DGS}(n)$ vanish  and do not contribute to the Euler-MacLaurin formula of Eq.~\eqref{EML}. 
We can thus apply also in this final example  the Euler-MacLaurin formula as given in  Eq.~\eqref{eq:EM} to calculate the Casimir energy, and we finally get:
\begin{eqnarray} 
  \mathcal{E}& = & -\frac{\pi^{2}}{720}\frac{\hbar c}{a^{3}}\left[1+\pi^{2}\frac{4\left(3+\pi^{2}\right)}{21}\left(\frac{\hbar\sqrt{\beta}}{a}\right)^{2}\right]
 \label{eq:RisultatoK}
\end{eqnarray}
 while the  Casimir pressure is given as:
\begin{eqnarray}
 \mathcal{F}= -\frac{\pi^{2}}{240}\frac{\hbar c}{a^{4}}\left[1+\pi^{2} \left(\frac{20}{21}+\frac{20\pi^2}{63}\right) \left(\frac{\hbar\sqrt{\beta}}{a}\right)^{2}\right]\, .\end{eqnarray}

\section{Discussion}
\label{discussion}
\begin{figure}
\scalebox{1.0}{\includegraphics{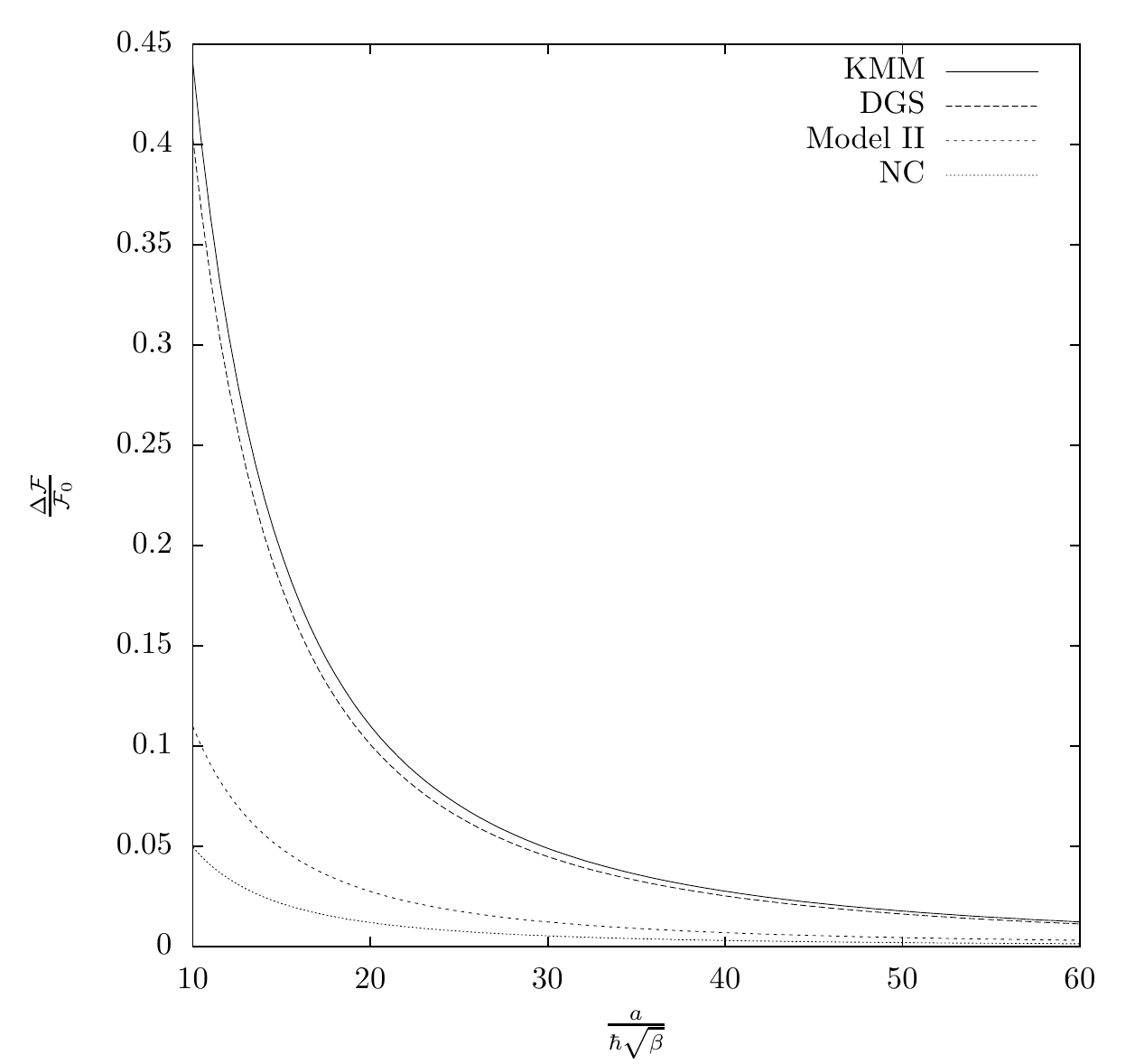}}
  \caption{ Plot of the correction terms for Model I with KMM maximally localized states (solid line) and DGS maximally localized states (dashed line); 
Model II (dot-dashed line) and in the noncommutative case Ref. \cite{PhysRevD.76.025016}  (dotted line). }
 \label{closurefig}
\end{figure}

For all  models  analyzed in this work the correction term has the same sign as the standard quantum field theory Casimir energy and the differences among various models are only in the values of the numerical constants.
Fig.~\ref{closurefig} shows the  differences of the Casimir pressure among the three models as a function of $a/\hbar \sqrt{\beta}$. 
There is a further dot-dashed line that is the result obtained in Ref \cite{PhysRevD.76.025016},  where the so called volume corrections  to the Casimir force  are due to space-time non commutativity. In Ref. \cite{PhysRevD.76.025016} the  coherent state approach leads to a nontrivial corrections already at the level 
of the free propagator as in Ref. \cite{PhysRevD.76.045012} and the corrections to the Casimir effect have the form of an attractive force.\\ 
The sign of the corrections to Casimir effect in theories that include a minimal length is a controversial issue (see Ref. \cite{2005JPhA...3810027N,2011arXiv1106.2737D} for repulsive corrections). 
It is interesting to note that in Ref. \cite{Altaisky:2011uc} the  correction to the Casimir energy is calculated  depending as a function the measurement resolution. 
The measurement resolution  is implemented by a real physical parameter  $\delta$   used to constrain maximal momenta of the field fluctuations and corrections 
are still attractive.

We can make some considerations about the possibility of observing
this effect. Clearly, should the minimal length $\Delta x_0 = \hbar \sqrt{\beta}$ be of the order of the Planck length, $L_p$,
no observation is possible. However, current experiments on the Casimir force can set an upper bound on the minimal length of the theory.   
In Ref.~\cite{PhysRevLett.88.041804}, the authors measure the coefficient of the Casimir force  between conducting surfaces in parallel configuration, 
with distance $a$ between the surfaces tested in the $0.5\, -\, 3.0 \, \mu m$ range and a precision of $15 \%$. 
Using this result, the upper bound obtained for the minimal plates distance $0.5 \, \mu m$ goes from  $ (\Delta x_0)_{\text{KMM}} = \hbar \sqrt{\beta} = 29 \, nm$ 
to $ (\Delta x_0)_{\text{II}}  = 58 \, nm$. As remarked at the beginning, parallel plates experimental configuration is more difficult then the sphere-plate geometry. 
This differences brings to a greater experimental accuracy \cite{PhysRevLett.81} and the calculation of Casimir corrections for non-planar geometries could give 
more stringent upper bounds on the minimal length than those reported here for the planar geometry (see Table~\ref{table1}).

\begin{table}\caption{\label{table1} Numerical upper bounds, in meters, on the 
minimal length as discussed in the text.
}
\begin{ruledtabular}
\begin{tabular}{lcccc}
$\,\,\,{\Delta x_0}$& $a$ & Model I (KMM) &Model I (DGS)  &  Model II \vspace{0.1cm}\\\hline
\vspace{0.1cm} 
$\hbar \sqrt{\beta}\,(\text{m})$&$3\cdot10^{-6}$&$ 1.750722362\cdot10^{-7}$& $ 1.750722362\cdot10^{-7}$& $3.508635606\cdot10^{-7}$\\
$\hbar \sqrt{\beta} \,(\text{m})$&$0.5\cdot10^{-6}$&$
 2.917870604\cdot10^{-8}
$&$3.049568826\cdot10^{-8}$&
$5.847726011\cdot10^{-8}$\vspace{0.2cm}\\
\end{tabular}
\end{ruledtabular}
\end{table}

\section{Conclusions}
\label{conclusion}
We have studied the correction to the Casimir energy given by models based on generalized Heisenberg uncertainty relations (GUP) that provide a minimal length. 
The configuration analyzed  has the geometry of  two parallel metallic plates with classical boundary conditions. Specifically, we have considered three models of quantum mechanics with a minimal length  $\Delta x_{0} = \hbar \sqrt{\beta}$ that differ among them by their physical states. A finite minimal uncertainty $\Delta x_{0}$ implies normalizzable maximal localization states, and thereby regularizes the ultraviolet region of the theories.
 The definition of these states is necessary to the quantization of the electromagnetic field. The field operators $A^i \left( \bm{r},t \right) $ instead of being 
expanded over a complete set of plane waves (position representation of momentum eigenfunctions) are now expanded in a complete set of \emph{maximally localized states}. 
The maximally localized states are just the physical states of the theory. This approach to the second quantization has been followed in Ref. \cite{PhysRevD.76.045012} to derive the Casimir-Polder intermolecular interactions in the presence of a minimal length. 
We then derive the Hamiltonian necessary to calculate the energy shift, known as the Casimir energy and hence the corrections to the Casimir energy due to a 
minimal length for all models separately. Because of  the natural cut-off induced by the theory, the Casimir energy does not need to be regularized as distinct from  the standard QED calculation.

We have decided to compute the leading order correction in $\beta$ to the Casimir energy in order to show an analytical result. This has been achieved by computing the boundary conditions  in Eq.~\ref{bc} at leading order in $\beta$ which is reflected in the fact that $n_{max} \to \infty$ and the Euler MacLaurin formula of Eq.~\eqref{EML} simplifies to that in Eq.~\eqref{eq:EM} because all contributing terms at $N=n_{\text{max}}\to +\infty$ vanish.

The first model  we consider (Model I) was proposed in Ref. \cite{1997PhRvD..55.7909K}, and has been the object of many phenomenological studies. 
Within this model, we distinguish between two different definitions of maximally localized states. The KMM procedure consists in minimizing the position 
uncertainty within the set of squeezed
states (see Ref.~\cite{kempf-1995-52},~\cite{1997PhRvD..55.7909K} for details), while the DGS procedure~\cite{2002PhRvD..66l5004D} is based on a 
minimization procedure on the subset of all physical states. 
For both these two definitions of maximally localized states we find an \emph{attractive correction} due to the minimal length to the Casimir potential energy.
 
 (Model II) is based on Ref.~\cite{2005JPhA...3810027N} where the Casimir energy was also calculated. As regards this point we might emphasize that the Casimir effect for parallel plates, calculated in Ref.  \cite{2005JPhA...3810027N} turns out to be repulsive, while in our computation the correction 
term is \emph{attractive} and therefore increases the (also attractive) force between two parallel plates due to ordinary QED. This result is in agreement with the analysis of the Casimir effect calculated using the 
proper maximally localized states (Model I) and is also in agreement  with the results discussed in~\cite{PhysRevD.76.045012} as regards the  minimal length correction to Casimir-Polder interactions. 

\pagebreak{}

\bibliography{CasimirDraft}
\bibliographystyle{unsrt}
\nocite{*}

\end{document}